\documentclass[12pt, prd, showpacs]{revtex4}
\usepackage{amssymb}
\usepackage{amsmath}

\input{tcilatex}

\begin{document}

\title{On geodesics with negative energies in the ergoregions of dirty black
holes}
\author{O. B. Zaslavskii}
\affiliation{Department of Physics and Technology, Kharkov V.N. Karazin National
University, 4 Svoboda Square, Kharkov 61022, Ukraine}
\affiliation{Institute of Mathematics and Mechanics, Kazan Federal University, 18
Kremlyovskaya St., Kazan 420008, Russia}
\email{zaslav@ukr.net }

\begin{abstract}
We consider behavior of equatorial geodesics with the negative energy in the
ergoregion of a generic rotating "dirty" (surrounded by matter) black hole.
It is shown that under very simple and generic conditions on the metric
coefficients, there are no such circular orbits. This entails that such
geodesic must originate and terminate under the event horizon. These results
generalize the observation made for the Kerr metric in A. A. Grib, Yu. V.
Pavlov, and V. D. Vertogradov, Mod. Phys.\ Lett. \textbf{29,} 1450110 (2014)
[arXiv:1304.7360].
\end{abstract}

\maketitle

\section{Introduction}

One of remarkable properties of rotating black holes consists in the
possibility of negative energies (measured with respect to infinity) inside
their ergosphere. This is the region where the metric coefficient $g_{00}>0$
(in the signature -, +, +, +). The properties of such trajectories are of
interest both from the theoretical and astrophysical viewpoints. The Penrose
effect \cite{pen} that leads to extraction of energy from a black hole is
intimately connected with such orbits. Therefore, it is important to know
the behavior of these trajectories in the vicinity of black hole. Recently,
an interesting observation was made in \cite{g3}. It turned out that in the
Kerr metric, there are no circular orbits of this kind. Therefore, all
orbits with the negative energy originate and terminate under the horizon.

Meanwhile, in real astrophysical conditions black holes are surrounded by
matter, so they are "dirty" in this sense. The goal of the present paper is
to formulate the conditions under which the aforementioned properties of the
geodesics under discussion are valid for a dirty rotating black hole. 

Throughout the paper we use units in which fundamental constants are $G=c=1$.

\section{Basic equations}

Let us consider the stationary axially symmetric metric of the form

\begin{equation}
ds^{2}=-N^{2}dt^{2}+R^{2}(d\phi -\omega dt)^{2}+\frac{dr^{2}}{A}+g_{\theta
}d\theta ^{2},  \label{met}
\end{equation}%
where all coefficients do not depend on $t$ and $\phi $. Let us consider
equations of motion for test geodesic particles. In the Kerr metric, the
variables in equations of motion can be separated \cite{car} and this
simplifies analysis greatly. In a general case, this is not so, and we
restrict ourselves by motion in the equatorial plane $\theta =\frac{\pi }{2}$
only. Then,%
\begin{equation}
m\dot{t}=\frac{X}{N^{2}}\text{,}  \label{t}
\end{equation}%
\begin{equation}
m\dot{\phi}=\frac{L}{R^{2}}+\frac{\omega X}{N^{2}}\text{,}  \label{fi}
\end{equation}%
where dot denotes derivative with respect to the proper time. Here, $m$ is
the mass,%
\begin{equation}
X=E-\omega L\text{,}  \label{x}
\end{equation}%
$E=-mu_{0}$ is the energy, $L=mu_{\phi }$ being the angular momentum, $%
u^{\mu }$ is the four-velocity. The quantities $E$ and $L$ are conserved due
to independence of the metric of $t$ and $\phi $, respectively. From the
normalization condition $u_{\mu }u^{\mu }=-1$, it follows that%
\begin{equation}
m\dot{r}=\pm \frac{\sqrt{A}}{N}Z\text{,}  \label{r}
\end{equation}

\begin{equation}
Z^{2}=X^{2}-N^{2}(m^{2}+\frac{L^{2}}{R^{2}})\text{.}  \label{z}
\end{equation}%
It can be rewritten as%
\begin{equation}
m^{2}\dot{r}^{2}+V_{eff}=0\text{, }V_{eff}=-\frac{A}{N^{2}}Z^{2}\text{.}
\label{v}
\end{equation}

By redefining a radial coordinate, one can always achieve that $A=N^{2}$ for
motion in the equatorial plane. We assume that this equality is satisfied.

The circular orbits $r=r_{0}=const$ are characterized by two equalities%
\begin{equation}
Z^{2}=0\text{,}  \label{circ1}
\end{equation}%
\begin{equation}
(Z^{2})^{\prime }=0,  \label{circ2}
\end{equation}%
prime denotes derivative with respect to $r$.

The forward in time condition $\dot{t}>0$ entails%
\begin{equation}
X\geq 0,  \label{x0}
\end{equation}%
where the equality sign is possible on the horizon $N=0$ only. We assume
that $\omega >0$ everywhere outside the horizon. We are interested in the
case $E<0$, so it follows from (\ref{x}) and (\ref{x0}) that 
\begin{equation}
L<0.  \label{L}
\end{equation}

Further, we assume that (like, say, for the Kerr metric),%
\begin{equation}
\omega ^{\prime }<0\text{,}  \label{om}
\end{equation}%
\begin{equation}
N^{\prime }\geq 0,  \label{N}
\end{equation}%
\begin{equation}
R^{\prime }\geq 0\text{.}  \label{R}
\end{equation}

\begin{theorem}
If inside the ergoregion%
\begin{equation}
(\omega R)^{\prime }<0,  \label{omR}
\end{equation}%
the circular orbits with the negative energy do not exist.
\end{theorem}

Now, we will prove this statement. It follows from (\ref{z}) that

\begin{equation}
\frac{1}{2}(Z^{2})^{\prime }=-X\omega ^{\prime }L-NN^{\prime }m^{2}-\frac{%
L^{2}}{2}(\frac{N^{2}}{R^{2}})^{\prime }  \label{z'}
\end{equation}

Let us suppose that the orbit is a circular one. Then, for such an orbit, (%
\ref{z}), (\ref{circ1}) entail 
\begin{equation}
X=N\sqrt{m^{2}+\frac{L^{2}}{R^{2}}}.
\end{equation}%
After substitution into (\ref{z'}), we obtain 
\begin{equation}
\frac{1}{2}(Z^{2})^{\prime }=-N\omega ^{\prime }L\sqrt{m^{2}+\frac{L^{2}}{%
R^{2}}}-NN^{\prime }m^{2}-\frac{L^{2}}{2}(\frac{N^{2}}{R^{2}})^{\prime }.
\end{equation}%
It follows from (\ref{L}) and (\ref{om}) that $\omega ^{\prime }L>0.$%
Therefore,%
\begin{equation}
\frac{1}{2}(Z^{2})^{\prime }\leq L^{2}[N\omega ^{\prime }\frac{1}{R}-\frac{1%
}{2}(\frac{N^{2}}{R^{2}})^{\prime }]=L^{2}\frac{N}{R}\chi ^{\prime }\text{,}
\end{equation}%
where%
\begin{equation}
\chi =\omega -\frac{N}{R}\text{.}
\end{equation}

It follows from (\ref{N}) that%
\begin{equation}
\chi ^{\prime }<\omega ^{\prime }+\frac{R^{\prime }N}{R^{2}}\text{.}
\end{equation}

It is seen from (\ref{met}) that

\begin{equation}
g_{00}=-N^{2}+R^{2}\omega ^{2}.
\end{equation}%
In the ergoregion, $g_{00}>0$, so%
\begin{equation}
N<\omega R\text{.}
\end{equation}%
Taking into account (\ref{R}), we have

\begin{equation}
\chi ^{\prime }<\omega ^{\prime }+\omega \frac{R^{\prime }}{R}=\frac{(\omega
R)^{\prime }}{R}\text{.}
\end{equation}

According to assumption (\ref{omR}), we have $(Z^{2})^{\prime }<0$ whereas
circular orbits require (\ref{circ2}, so they are impossible. Thus the
statement is proven.

We can somewhat modify and simplify condition (\ref{omR}) replacing it
directly with%
\begin{equation}
\chi ^{\prime }<0\text{.}  \label{xneg}
\end{equation}%
As $\chi >0$ near the horizon and $\chi =0$ on the boundary of the
ergoregion, (\ref{xneg}) is equivalent to the condition that inside the
ergosphere the function $\chi $ is monotonically decreasing.

\section{Example: Kerr-Newman metric}

In this section, we consider the Kerr-Newman metric and show that the
criteria (\ref{om}) - (\ref{omR}) are satisfied. Taking the expression for
the metric from any textbook, one can find that for $\theta =\frac{\pi }{2}$%
, 
\begin{equation}
R^{2}=r^{2}+a^{2}+\frac{(2Mr-Q^{2})a^{2}}{r^{2}},
\end{equation}%
\begin{equation}
N^{2}=\frac{r^{2}-2Mr+Q^{2}+a^{2}}{R^{2}}\text{,}
\end{equation}

\begin{equation}
\omega =\frac{a(2Mr-Q^{2})}{R^{2}r^{2}}\equiv \frac{aP}{R^{2}}\text{, }P=%
\frac{2M}{r}-\frac{Q^{2}}{r^{2}}.  \label{oma}
\end{equation}%
Here, $r$ is a Boyer-Lindquiste coordinate, $M$ is the black hole mass, $a=%
\frac{J}{M}$, $J$ being its angular momentum, $Q$ electric charge.

The event horizon lies at%
\begin{equation}
r=r_{+}=M+\sqrt{M^{2}-Q^{2}-a^{2}}\text{, }M^{2}\geq a^{2}+Q^{2}\text{.}
\label{hor}
\end{equation}%
Then,%
\begin{equation}
(R^{2})^{\prime }=\frac{2}{r^{3}}[r(r^{3}-a^{2}M)+a^{2}Q^{2}].  \label{r'}
\end{equation}

As%
\begin{equation}
r>M  \label{rm}
\end{equation}%
outside the horizon, both terms in (\ref{r'}) are positive, eq. (\ref{R}) is
satisfied.

It is easy to check that outside the horizon, 
\begin{equation}
P>0,\text{ }P^{\prime }<0  \label{p}
\end{equation}
Then, it follows from (\ref{oma}) and (\ref{R}) that conditions (\ref{om})
and (\ref{omR}) are satisfied.

One can also obtain that%
\begin{equation}
(N^{2})^{\prime }=\frac{2r(S_{1}+S_{2})}{(Ar^{2}+Ba^{2})^{2}}\text{, }%
A=r^{2}+a^{2}\text{, }B=2Mr-Q^{2}\text{.}
\end{equation}%
\begin{equation}
S_{1}=Ar(Mr^{2}-rQ^{2}-Ma^{2})\text{, }S_{2}=Ba^{2}(r^{2}-2Mr+a^{2}+Q^{2})%
\text{.}
\end{equation}%
It is obvious that $S_{2}>0$ outside the horizon. We can write%
\begin{equation}
Mr^{2}-rQ^{2}-Ma^{2}=r(Mr-Q^{2})-Ma^{2}\geq r(M^{2}-Q^{2})-Ma^{2}\geq
M(M^{2}-Q^{2}-a^{2})\geq 0\text{.}
\end{equation}%
Therefore, $S_{1}>0$ as well, so condition (\ref{N}) is satisfied.

Thus we checked all eqs. (\ref{om}) - (\ref{omR}) for the Kerr-Newman
metric. For $Q=0$ we return to the Kerr metric, the results are in agreement
with \cite{g3}.

\section{Conclusion}

Conditions (\ref{om}) - (\ref{R}) are very simple and reasonable. They mean
that the properties of the metric coefficients from the horizon to infinity
change monotonically. The rotation parameter $\omega $ coincides with the
angular velocity of a black hole on the horizon and tends to zero far from a
black hole. The lapse function $N=0$ on the horizon and $N\rightarrow 1$ at
infinity for asymptotically flat space-times. The coefficient $R^{2}$
determines the surface area of cross-sections $t=const,$ $r=const$, $\theta
=const$ and increases from the horizon to infinity (absence of wormholes),
if (\ref{R}) is satisfied. Thus there is nothing specific in conditions
under discussion. Eq. (\ref{omR}) is slightly more special but it means only
that $\omega $ should decrease more rapidly than $1/R$.

Therefore, one can conclude that the properties of geodesics found in \cite%
{g3} for the Kerr metric, are quite generic. Thus the orbits with negative
energies cannot originate and terminate in the ergoregion outside the
horizon. Correspondingly, this can happen under the event horizon only.

Consideration in the present paper applies to equatorial geodesics only. It
would be interesting to elucidate, whether this general approach can be
generalized to nonequatorial motion.


\begin{thebibliography}{9}
\bibitem{pen} R. Penrose, Riv. Nuovo Cimento 1, 252 (1969); reprinted in
Gen. Relativ. Gravit. \textbf{34}, 1141 (2002).

\bibitem{g3} A. A. Grib, Yu. V. Pavlov, and V. D. Vertogradov, Mod. Phys.\
Lett. \textbf{29,} 1450110 (2014) [arXiv:1304.7360].

\bibitem{car} B. Carter, Phys. Rev. \textbf{174}, 1559 (1968).
\end{thebibliography}
\end{document}